\documentclass[twocolumn]{aastex631}

\usepackage[utf8]{inputenc}
\usepackage[english]{babel}
\usepackage{natbib}
\usepackage{fp}
\usepackage[capbesideposition=right]{floatrow}
\usepackage{wrapfig}
\usepackage{mathtools}
\usepackage{amsmath}
\usepackage{supertabular,enumitem}
\usepackage{tabularx}
\usepackage{float}

\submitjournal{ApJ}

\shorttitle{Connecting the Low to High Corona}
\shortauthors{Alzate et al.}
\begin{document}

\title{Connecting the Low to High Corona:  A Method to Isolate Transients in \textit{STEREO}/COR1 Images}

%%%%%%%%%%%%%%%%%%%%%%%%

\correspondingauthor{Nathalia Alzate}
\email{nathalia.alzate@nasa.gov}

\author[0000-0001-5207-9628]{Nathalia Alzate}
\affiliation{NASA Goddard Space Flight Center \\
Greenbelt MD 20771, USA}
\affiliation{ADNET Systems, Inc. \\
Greenbelt MD 20771, USA}

\author[0000-0002-6547-5838]{Huw Morgan}
\affiliation{Aberystwyth University \\
Ceredigion, Cymru SY23 3BZ, UK}

%Institute of Mathematics, Physics and Computer Science, Aberystwyth University, Ceredigion, Cymru, SY23 3BZ, UK

\author[0000-0002-8767-7182]{Nicholeen Viall}
\affiliation{NASA Goddard Space Flight Center \\
Greenbelt MD 20771, USA}

\author[0000-0002-8164-5948]{Angelos Vourlidas}
\affiliation{Johns Hopkins University Applied Physics Laboratory \\
Laurel, MD 20723, USA}

%% Mark off the abstract in the ``abstract'' environment. 
\begin{abstract}

We present a method that isolates time-varying components from coronagraph and EUV images, allowing sub-streamer transients propagating within streamers to be tracked from the low to high corona. The method uses a temporal bandpass filter with a transmission bandwidth of $\sim$2.5-10 hours that suppresses both high and low frequency variations in observations made by the STEREO/SECCHI suite. We demonstrate that this method proves crucial in linking the low corona where the magnetic field is highly non-radial, to their counterparts in the high corona where the magnetic field follows a radial path through the COR1 instrument. We also applied our method to observations by the COR2 and EUVI instruments onboard SECCHI and produced height-time profiles that revealed small density enhancements, associated with helmet streamers, propagating from $\sim$1.2 $R_{\odot}$ out to beyond 5 $R_{\odot}$.  Our processing method reveals that these features are common during the period of solar minimum in this study. The features recur on timescales of hours, originate very close to the Sun, and remain coherent out into interplanetary space. We measure the speed of the features and classify them as: slow (a few to tens of $kms^{-1}$) and fast ($\sim$100 $kms^{-1}$). Both types of features serve as an observable tracer of a variable component of the slow solar wind to its source regions. Our methodology helps overcome the difficulties in tracking small-scale features through COR1. As a result, it proved successful in measuring the connectivity between the low and high corona and in measuring velocities of small-scale features.

\end{abstract}

%% Keywords should appear after the \end{abstract} command. 
%% See the online documentation for the full list of available subject
%% keywords and the rules for their use.


%% From the front matter, we move on to the body of the paper.
%% Sections are demarcated by \section and \subsection, respectively.
%% Observe the use of the LaTeX \label
%% command after the \subsection to give a symbolic KEY to the
%% subsection for cross-referencing in a \ref command.
%% You can use LaTeX's \ref and \label commands to keep track of
%% cross-references to sections, equations, tables, and figures.
%% That way, if you change the order of any elements, LaTeX will
%% automatically renumber them.
%%
%% We recommend that authors also use the natbib \citep
%% and \citet commands to identify citations.  The citations are
%% tied to the reference list via symbolic KEYs. The KEY corresponds
%% to the KEY in the \bibitem in the reference list below. 

%%%%%%%%%%%%%%%%%%%%%%%%%%%%%%%%%%%%%%%
%%%  	 INTRODUCTION
%%%%%%%%%%%%%%%%%%%%%%%%%%%%%%%%%%%%%%%

\section{Introduction} 
\label{sec:intro}

Establishing a direct connection between the low and the high corona is a crucial aspect of understanding how the Sun forms the heliosphere.  In doing so, the origin and evolution of solar transient structures, and their subsequent impact on the heliosphere, can be determined.  Further, establishing this connectivity is a crucial first step for identifying the sources of the slow solar wind.  As of yet, there is no fully-understood direct link between structures observed low in the corona and those higher up in and near streamers.

In general, it is known that the slow, highly variable and dense solar wind is associated with the streamer belt \citep{Hundhaussen_1977, Gosling_1981, McComas_1998}.  The streamer belts exhibit a high variation in density on smaller spatial scales \citep{Morgan_2020, Decraemer_2019}.  The streamers also show considerable temporal variation of density, and are host to several small-scale dynamic features \citep{Sheeley_1997, Jones_2009, Diaz_2017, DeForest_2018}.  Observations by heliospheric imagers and coronagraphs reveal that structures in the high corona become structures in the solar wind.  However, linking those structures to dynamics and structures (faint and small ones in particular) in the low corona, is uncertain below $\sim$2.5 $R_{\odot}$.  This is because plasma motion exhibits non-radial flows due to the highly non-radial structure of the magnetic field.

Improving the understanding of the formation of small-scale dynamic features requires continuous observations over the distance range starting from the inner boundary of the corona extending out to several solar radii.  Typically, this is achieved by trying to link extreme ultraviolet (EUV) observations in the low corona with white light (WL) observations in the mid and upper corona.  Extreme ultraviolet and WL observations are difficult to interpret due to the line-of-sight effect and instrumental limitations.  At present, without a total solar eclipse (TSE), there is not a single dataset that spans an uninterrupted field of view (FOV) of this distance range.  Since TSE observations are done over short periods of time, current observations of the low corona thus fall short in placing the dynamic behavior of small-scale structures very low in the corona within the context of the high corona.   Combining results from multiple instruments is therefore required to capture this dynamic connectivity.  The FOV of the COR1 inner coronagraph \citep{Thompson_2003} on \textit{Sun Earth Connection Coronal and Heliospheric Investigations} \citep[SECCHI, ][]{Howard_2008} onboard the \textit{Solar TErrestrial RElations Observatory Ahead and Behind} \citep[STEREO A \& B, ][]{Kaiser_2008} twin spacecraft, extends from $\sim$1.4 out to 4.0 $R_{\odot}$.  The SECCHI-COR2 outer coronagraph instruments observe the corona between $\sim$2.5 and 15 $R_{\odot}$.  Data from the Extreme Ultraviolet Imagers (EUVI) on STEREO observe the corona up to $\sim$1.7 $R_{\odot}$.  Together, with careful processing, these observations enable a continuous view of the extended corona.  However, due to the difficulty in observing the white light corona close to the Sun, COR1 data present a challenge to analysis of small-scale structures.

% FIGURE 1 %%%%%%%%%%%
% cor1 difference
\begin{figure}
\centering
%\begin{interactive}{animation}{fig1anim.mp4}
\includegraphics[width=0.9\textwidth]{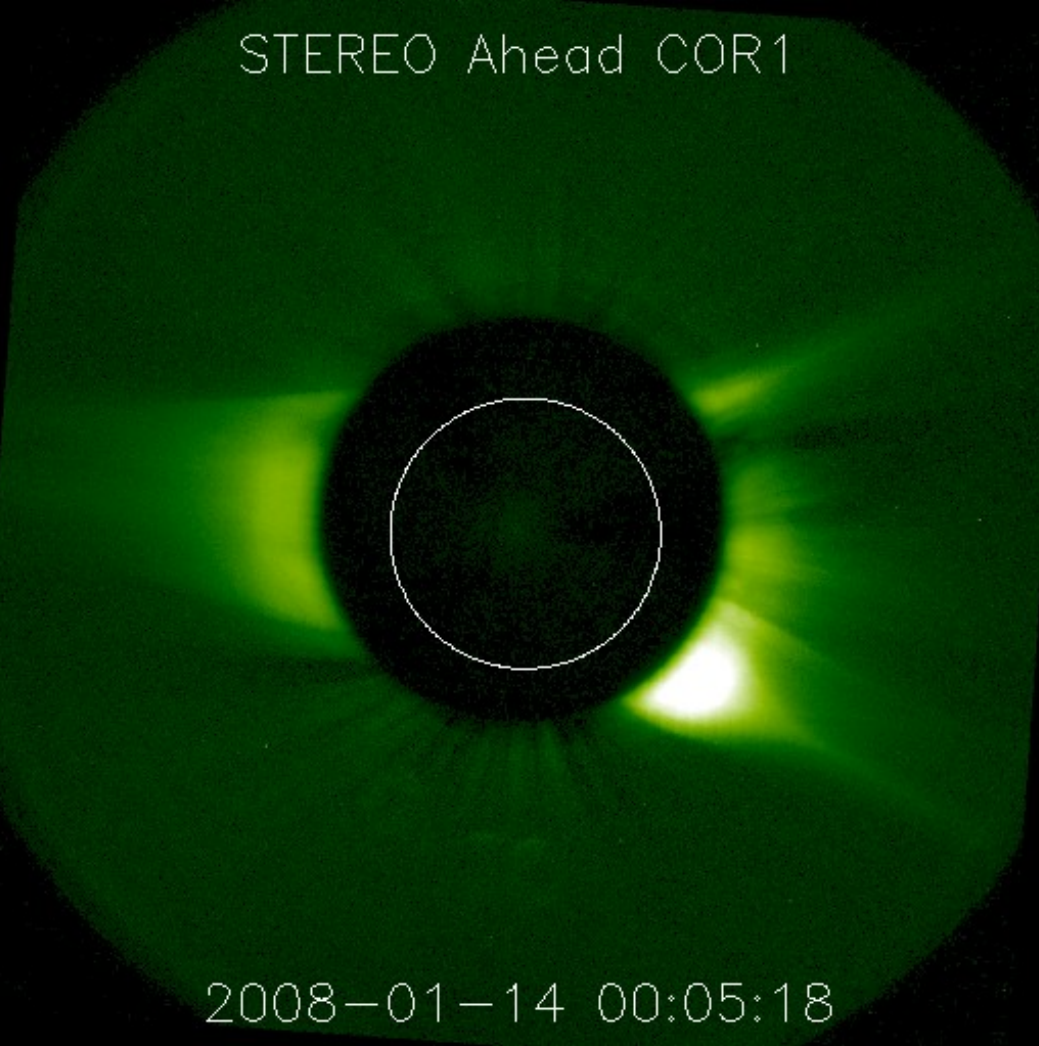}
%\end{interactive}
\caption{\textit{STEREO}/COR1 background-subtracted image \citep{Thompson_2010} from  the \textit{STEREO} Science Center website (\url{https://stereo-ssc.nascom.nasa.gov/}).  An animated version is provided online. The animation spans 13-17 January 2008 at a rate of 5 minutes per frame (41 seconds total duration). \label{fig:diff_cor1}}
\end{figure}
%%%%%%%%%%%%%%%%%%

A common approach for deriving useful data from light-scattered signals is to correct for the background to reveal the underlying K-corona, B$_{k}$, or electrons.  This can be done via the routine {\fontfamily{cmtt}\selectfont SECCHI\_PREP} within the \textit{Interactive Data Language} (IDL) based system, \textit{SolarSoft} \citep{Freeland_1998}.  A COR1 movie of background-corrected images is available as part of the online version of this article.  We generated it using the monthly minimum background-corrected \citep[see][]{Thompson_2010} images provided online via the \textit{STEREO} Science Center website\footnote{\url{https://stereo-ssc.nascom.nasa.gov/}}.  The movie shows an obvious increase in brightness of the streamer near the equator on the east limb, followed by an expulsion of plasma material that crosses the entire FOV.  The data used for this movie is dominated by scattered light.  Without higher level processing applied, only bright, large and slow-changing structures are revealed, leaving much of the sub-streamer structures hidden underneath the global brightness structure.  A still frame is shown in Figure \ref{fig:diff_cor1}.

Enhancing faint details in images of the low corona is required for tracking small-scale transients from near the Sun outwards. Previous studies by \citet{Alzate_2016, Alzate_2017} have focused on the application of advanced image processing techniques, namely the Normalized Radial Graded Filter \citep[NRGF, ][]{Morgan_2006}, the Dynamic Separation Technique \citep[DST, ][]{Morgan_2012}, and the Multiscale Gaussian Normalization Technique \citep[MGN, ][]{Morgan_2014} to coronal imaging data from EUV and coronagraph instruments.  However, the noise in COR1 data remains a challenge unsuitable for the DST and MGN (see corresponding references for details).

In this paper, we present our method for processing COR1 data that uses a temporal bandpass filter to enhance faint and small-scale (sub-streamer) transients.  To test our method, we applied it to COR1 observations over a period of relatively quiet Sun.  We further demonstrate the method as applied to COR2 and EUVI data in order to track features as they propagate through the three FOVs.  In Section \ref{sec:obs-meth} we describe the datasets used and present our method in detail as it applies to COR1, with the extension to EUVI and COR2.  Section \ref{sec:anal-res} presents the results from successfully testing our method on data for 10-23 January 2008.  In Section \ref{sec:discuss} we discuss our interpretation of the results.  In Section \ref{sec:conclusions} we summarize our work and conclude with implications for future studies of the slow solar wind.

%%%%%%%%%%%%%%%%%%%%%%%%%%%%%%%%%%%%%%%
%%%   			OBSERVATIONS & METHODS
%%%%%%%%%%%%%%%%%%%%%%%%%%%%%%%%%%%%%%%

\section{Observations \& Method} 
\label{sec:obs-meth}

%%%%%%%%%%%%%
\subsection{Observations and Pre-Processing} 
\label{sec:observations}
%%%%%%%%%%%%%

For this study, we chose a period of 14 days of low solar coronal activity during 10-23 January 2008.  We used data from the EUVI instrument and from the COR1 and COR2 white light coronagraphs onboard the \textit{STEREO A} spacecraft.  We used images in the 171 \AA\ (dominated by Fe IX ion emission with a peak formation temperature of 0.8 million K) and 195 \AA\ (dominated by Fe XII ion emission with a peak formation temperature of 1.4 million K) channels for EUVI.  We used polarized brightness observations of COR1 and COR2, which are taken in sequences of three images at polarization angles of 0$^{\circ}$, 120$^{\circ}$ and 240$^{\circ}$.  We used the largest image size available from COR1 observations during the 14-day period, which is 1024 $\times$ 1024 pixels (rebinned from the native 2048 $\times$ 2048 pixels onboard the instrument), at a 5-minute observation cadence.  The COR2 observations we used have a 30-minute cadence. 

For COR1 and COR2 observations, pre-processing started with applying the instrumental calibration via the standard SolarSoft {\fontfamily{cmtt}\selectfont SECCHI\_PREP} \textit{IDL} procedure.  This procedure calibrates and combines the three polarization images into an estimate of polarized brightness (pB), in units of mean solar brightness, when the keywords {\fontfamily{cmtt}\selectfont /polariz\_on} and  {\fontfamily{cmtt}\selectfont /pb} are used (see \citet{Thompson_2010} for details).  Then we applied a spatial point filter to each pB image to remove isolated pixels that have spurious high or low values compared to neighboring pixels.  The point filter iteratively removes outlying pixels by comparing brightness values at each pixel with local means and standard deviations.  For each pixel, a neighborhood of 5 $\times$ 5 pixels was used to calculate the local mean and standard deviation.  Pixels with a value deviating more than three standard deviations from the local mean were deemed as outlying and replaced by the local mean.  The process was then repeated until no outliers were identified.  

For EUVI observations, we extracted files with similar exposure times resulting in an irregular time series.  In order to have data at a fixed cadence, necessary for the subsequent steps, we interpolated the irregular time series onto a regular time grid with an increment of 18.5 minutes corresponding to the largest step we found in the data.  We then applied the same spatial point filter to the interpolated time series.

%%%%%%%%%%%%%%%%%%%%%%%%%%%%%%%%%%%%%%%%%%%%%%%%
\subsection{Method for Generating STEREO/COR1 Images and Application to COR2 and EUVI}
\label{sec:meth-img}
%%%%%%%%%%%%%%%%%%%%%%%%%%%%%%%%%%%%%%%%%%%%%%%%

The core of the processing method is a bandpass filter that operates in the temporal domain to effectively damp high-frequency noise and low-frequency slow-changing structures.  The filtering is achieved through convolution with two normalized kernels defined as a wide and a narrow Gaussian kernel.  The wide Gaussian kernel isolates and removes low frequency variations (e.g. the slow-varying structures), while the narrow Gaussian kernel suppresses high-frequency variations (e.g. noise).  We applied the method to the time series extracted from each pixel in the series of images in the following order:

\begin{enumerate}[itemsep=0mm]
  \item Create a wide normalized Gaussian kernel, $k_{w}$, of width $w_{w}$
  \item Smooth the time series of observations with the wide kernel $k_{w}$
  \item Subtract the smoothed profile from the original signal profile
  \item Create a narrow normalized Gaussian kernel, $k_{n}$, of width $w_{n}$
  \item Smooth the resulting time series with the narrow kernel $k_{w}$
\end{enumerate}

The Gaussian kernels are created via the {\fontfamily{cmtt}\selectfont GAUSSIAN\_FUNCTION} function in \textit{IDL}.  The parameter $\sigma$ in the function is the standard deviation and determines the width of the Gaussian.  The datasets vary in their temporal sampling rates, therefore we adjust the filter parameters in our method such that the effective bandpass filter is similar.  Table \ref{table:params} lists the observation cadences for each dataset, the parameters used and the corresponding time scales.  For this study, the wide and narrow Gaussian kernel's $\sigma$ are chosen to isolate timescales between $\sim$2.5 and $\sim$10 hours ($t_{n}$ and $t_{w}$).

%%%  Table with parameters used for each dataset:
%
\begin{table}
%\centering
\setlength{\tabcolsep}{4pt}
\begin{tabular}{|c|c|c|c|c|c|c|c|}
\hline
 \it{STEREO}    &    Cadence   &                  \multicolumn{3}{c|}{Narrow Filter}           &                \multicolumn{3}{c|}{Wide Filter}                \\ \cline{3-8}
%\hline
        Data          &       (min)      &  $\sigma_{n}$    &      $w_{n}$       &   $t_{n}$ (hr)   &   $\sigma_{w}$    &      $w_{w}$    &    $t_{w}$ (hr)   \\
 \hline\hline
        EUVI         &       18.5       &           1               &            7            &      2.16          &            5               &           31         &      9.56          \\
        COR1       &         5           &          5               &           31           &      2.58          &           20              &          123        &     10.25          \\
        COR2       &        30          &         0.4             &            7            &      3.50          &            3               &           19         &       9.50          \\
\hline
\end{tabular}
\caption{Parameters for each dataset.  The parameter $\sigma$ is the standard deviation in the Gaussian function that determines the width $w$ of the Gaussian corresponding to time $t$.  The subscripts `$w$' and `$n$' relate to the wide and narrow kernels, respectively.  Also listed are the cadences for each dataset.}  
\label{table:params}
\end{table}

\begin{figure}
\centering
\includegraphics[width=0.935\textwidth]{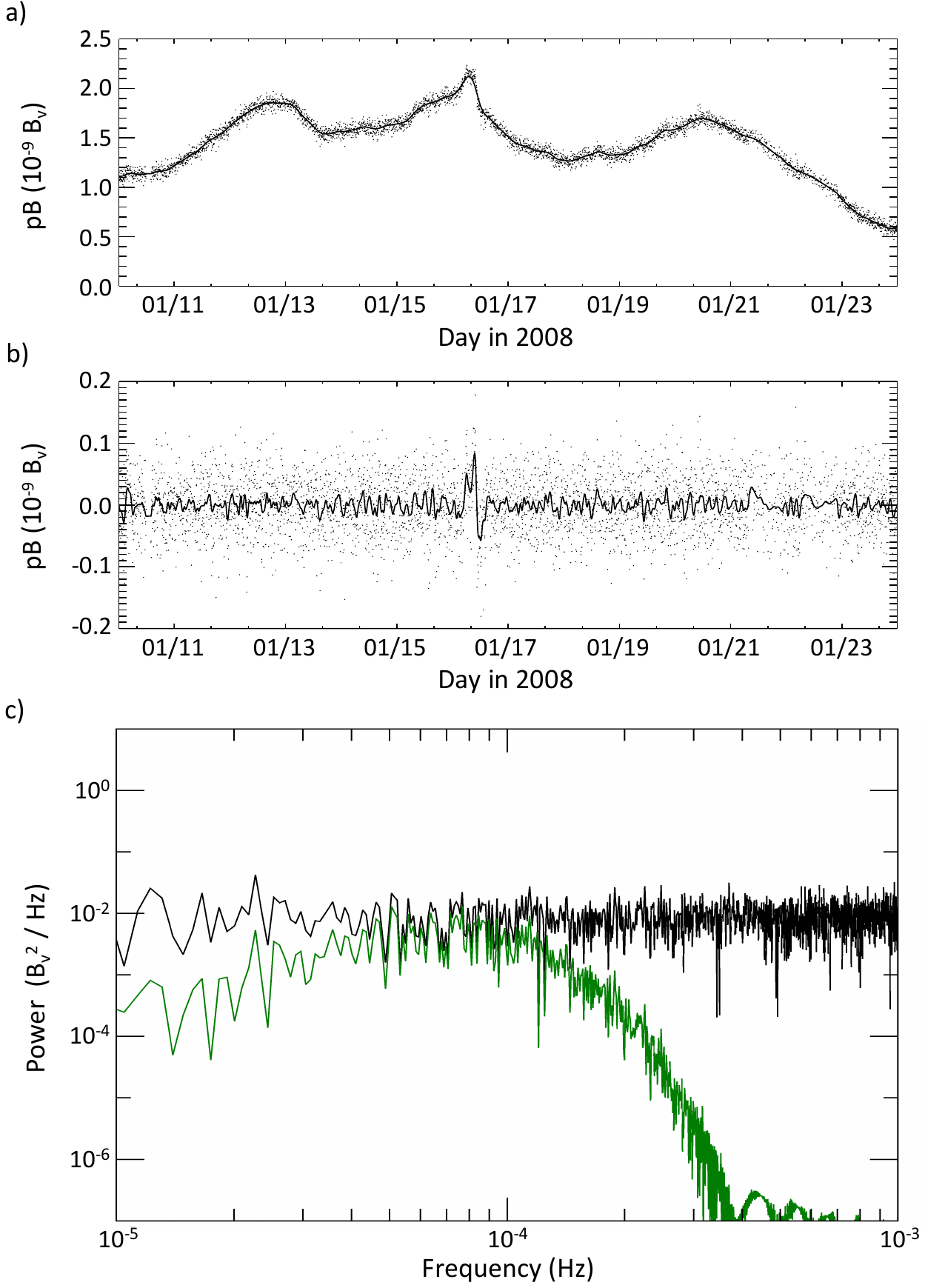}
\caption{a) Polarized brightness (pB) time series for a pixel at position angle 80$^{\circ}$ and a height of $\sim$2.6 $R_{\odot}$ represented by the dots.  The solid line represents the result from the wide Gaussian convolution.  b) The points are from subtracting the wide kernel result from the original signal.  Convolving this signal with the narrow Gaussian gives the result represented by the solid line, which is the final filtered signal.  In the example we show here, the high-amplitude disturbance during 16 January 2008 is a small CME.  c) The power spectrum of the original signal is shown in black and the power spectrum of the filtered signal is shown in green.   \label{fig:ts80}}
\end{figure}

Figure \ref{fig:ts80} is composed of three plots that display the different stages described in this section when applied to a time series of COR1 observations.  The dots in Figure \ref{fig:ts80}a show the pB time series for a pixel at a heliocentric height of 2.6 $R_{\odot}$ and position angle of 80$^{\circ}$.  Smoothing the time series at each pixel with the wide Gaussian kernel results in the profile shown as a solid line.  Subtracting this smoothed profile from the original signal (high pass filtering) gives the dotted profile in panel b).  The resulting time series is then smoothed with the narrow Gaussian kernel (low pass filter) to obtain the profile shown as a solid line.  Panel c) compares the power spectrum of the original time series (black) and the spectrum of the filtered time series (green) in the 10$^{-5}$ to 10$^{-3}$ Hz range.  Components on timescales longer than $\sim$10 hours and shorter than $\sim$2.5 hours are damped by more than one order of magnitude.  Note that any slow-changing or static signal was damped by the low bandpass filter.

% FIGURE 2 %%%%%%%%%%%
% Time series for one pixel
%%%%%%%%%%%%%%%%%%

%%%%%%%%%%%%%%%%%%%%%%%%%%%%%%%%%%%%%
\subsection{Application:  COR1 Images}
\label{sec:app-meth-img}
%%%%%%%%%%%%%%%%%%%%%%%%%%%%%%%%%%%%%

% FIGURE 3 %%%%%%%%%%%%%%%%
\begin{figure}
\centering
%\begin{interactive}{animation}{fig3anim.mp4}
\includegraphics[width=0.775\textwidth]{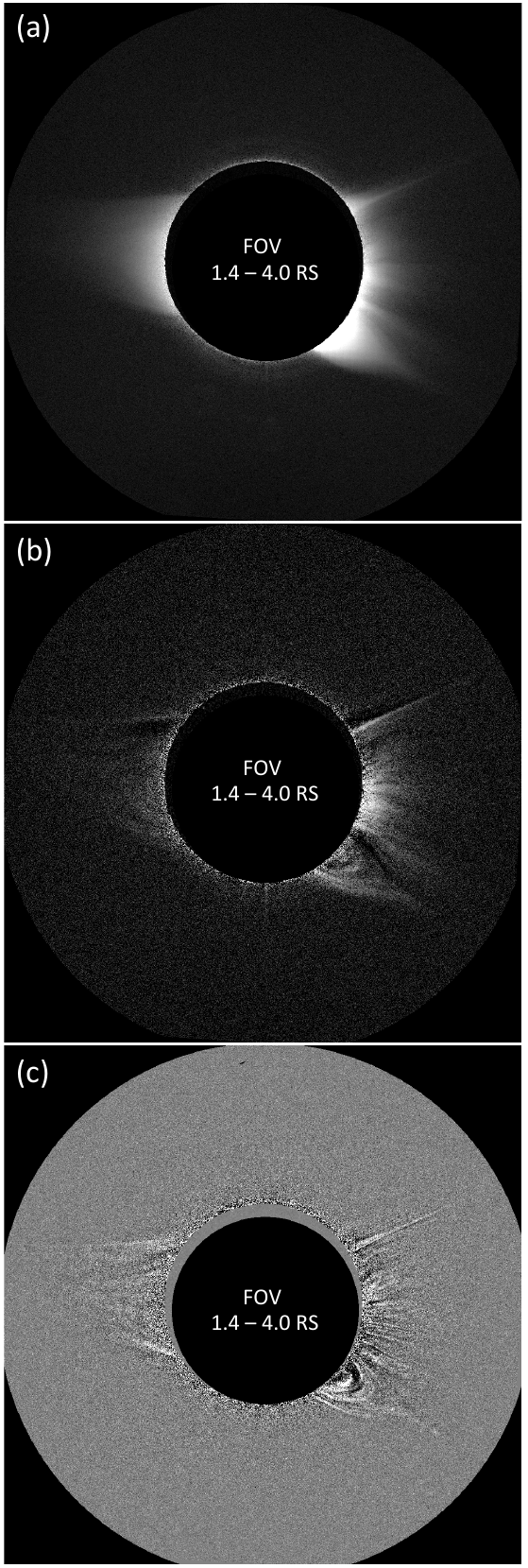}
%\end{interactive}
\caption{STEREO A/COR1 2008 January 14 observation at 03:10 UT at various steps during the filtering process.  (a) Unprocessed pB image with a point filter aplied.  (b)  Image following convolution with the wide Gaussian kernel subtracted from (a).  c)  Image following convolution with the narrow Gaussian kernel.  A movie of processed (filtered) COR1 images spanning 5 of the 14 days of observations (13-17 January) used in this study is available online.  \label{fig:bff-fulldisk}}
\end{figure}
%%%%%%%%%%%%%%%%%%%%%%%

For spatial context, we applied our filtering method, as described in Section  \ref{sec:meth-img}, to the 14-day time series of COR1 observations.    Figure \ref{fig:bff-fulldisk} shows the results at various steps of the processing, for still images of 14 January 2008 at 03:10 UT.  Applying a point filter to the (unprocessed) image of estimated total brightness (pB) results in the image shown in panel (a) where several bright, large-scale structures are visible.  The next step of applying the wide Gaussian kernel to the COR1 time series and subtracting the results from the time series of point-filtered images yields the image shown in panel (b).  Here, we are able to see substructures that were hidden against the brightness of the global scale structures in (a).  Panel (c) shows the result following the convolution with the narrow Gaussian kernel to the wide-Gaussian filtered COR1 time series.  This step removes structures on timescales below $\sim$2.5 hours.  The high contrast in brightness across the image in (b) is adjusted during this final step where the remaining brightness values are rescaled across the image (via the {\fontfamily{cmtt}\selectfont HIST\_EQUAL} function in IDL).  The finer sub-structures that were otherwise hidden within the brightness of the streamer in (a) and (b) are enhanced.  Figure \ref{fig:bff-fulldisk} demonstrates how our filtering method is able to reveal small-scale features in COR1 on the timescales defined in the bandwidth filter.  A movie of full disk COR1 bandpass-filtered images where the evolution and propagation of small-scale features are best seen is provided online.

%%%%%%%%%%%%%%%%%%%%%%%%%%%
\subsection{Method for Generating Height-Time Plots}
\label{sec:meth-plot}
%%%%%%%%%%%%%%%%%%%%%%%%%%%

We generated height-time (Ht-T) plots in order to track features as they propagate in time.  We constructed them from radial slices centered along a position angle (PA) from a succession of images.  To begin, we converted the point-filtered pB images (Section \ref{sec:observations}) from image pixel coordinates into heliocentric polar coordinates.  We then rebinned the polar images into number of height and number of position angle gins combining multiple pixels to reduce noise.  We then applied the filtering method to the rebinned images (datacube) following the steps in Section \ref{sec:meth-img} resulting in a datacube of filtered time series.  We extracted the radial slices of data at selected angle bins to generate the Ht-T plots.  The steps of our procedure are:

\begin{enumerate}[itemsep=0mm]
  \item Convert the point-filtered pB images from pixel coordinates into polar coordinates
  \item Define number of height bins, $n_{ht}$, within a FOV of $R_{high} - R_{low}$ in solar radii increments of $\Delta_{ht}=(R_{high} - R_{low})/n_{ht}$
  \item Define number of position angle bins, $n_{pa}$
  \item Define the width of the position angle bin, $\Delta _{pa}$, which can be chosen to be equal, or greater than, the angle bin step, that is 360$^{\circ}/n_{pa}$
  \item Re-bin the polar coordinate pB images into $n_{ht}$, $n_{pa}$ and time bins, $n_{t}$, to average values of the original image pixels that lie within $\Delta_{ht}$, and within $\Delta _{pa}$
  \item Apply the Gaussian kernels  to the time series at each binned polar-coordinate individually using the steps in Section \ref{sec:meth-img}.
  \item For display, at a given height in the datacube, the brightness is divided by the standard deviation of brightness at that height (calculated across all position angles and time)
\end{enumerate}

Following Step 5, there could be a small overlap between adjacent position angle bins (equivalent to smoothing along position angle) according to the choice of $\Delta _{pa}$.  Wide position angle bins, which help reduce noise variance, are particularly useful for COR1 where the noise level is high.  Furthermore, with increasing height (and decreasing signal), the position angle extent encompasses an increasing number of pixels, which, in turn, increases signal-to-noise ratio.  Step 7 is carried out for display purposes to effectively reveal the continuation of moving structure since the signal decreases rapidly with increasing height.  Table \ref{table:params-htt} lists the datacube dimensions, the height bin FOV range and the width of the radial slices for each dataset used in this study.

\begin{table}
%\centering
\setlength{\tabcolsep}{8pt}
\begin{tabular}{| c | c | c | c | }
\hline
 \it{STEREO}   &   $\Delta pa$    &                      Dimensions                       &    Field of View        \\ 
        Data        &     ($^{\circ}$)    &    $n_{pa} \times n_{ht} \times n_{t}$     &       ($R_{s}$)     \\
 \hline\hline
        EUVI       &            5            &           $72 \times 100 \times 1089$        &            1.0 to 1.6              \\
        COR1     &            5             &           $72 \times 100 \times 4032$       &             1.4 to 4.0              \\
        COR2     &            5             &            $72 \times 100 \times 675$        &             3.0 to 15.0            \\
\hline
\end{tabular}
\caption{Datacube parameters for each dataset:  $n_{pa}$ is the number of position angle bins, $n_{ht}$ is the number of height bins, $n_{t}$ is the number of time bins in the time series, and  $\Delta pa$ is the width of the radial slice.  Each dataset cadence determines $n_{t}$.  Also listed are the height bin FOV ranges.}
\label{table:params-htt}
\end{table}

%%%%%%%%%%%%%%%%%%%%%%%%%%%%%%%%%%%%%
\subsection{Application:  COR1 Height-Time Plots}
\label{sec:app-meth-plot}
%%%%%%%%%%%%%%%%%%%%%%%%%%%%%%%%%%%%%

% FIGURE 4 %%%%%%%%%%%%%%%%
%%bff-steps-fulldisk
\begin{figure}
\includegraphics[width=\textwidth]{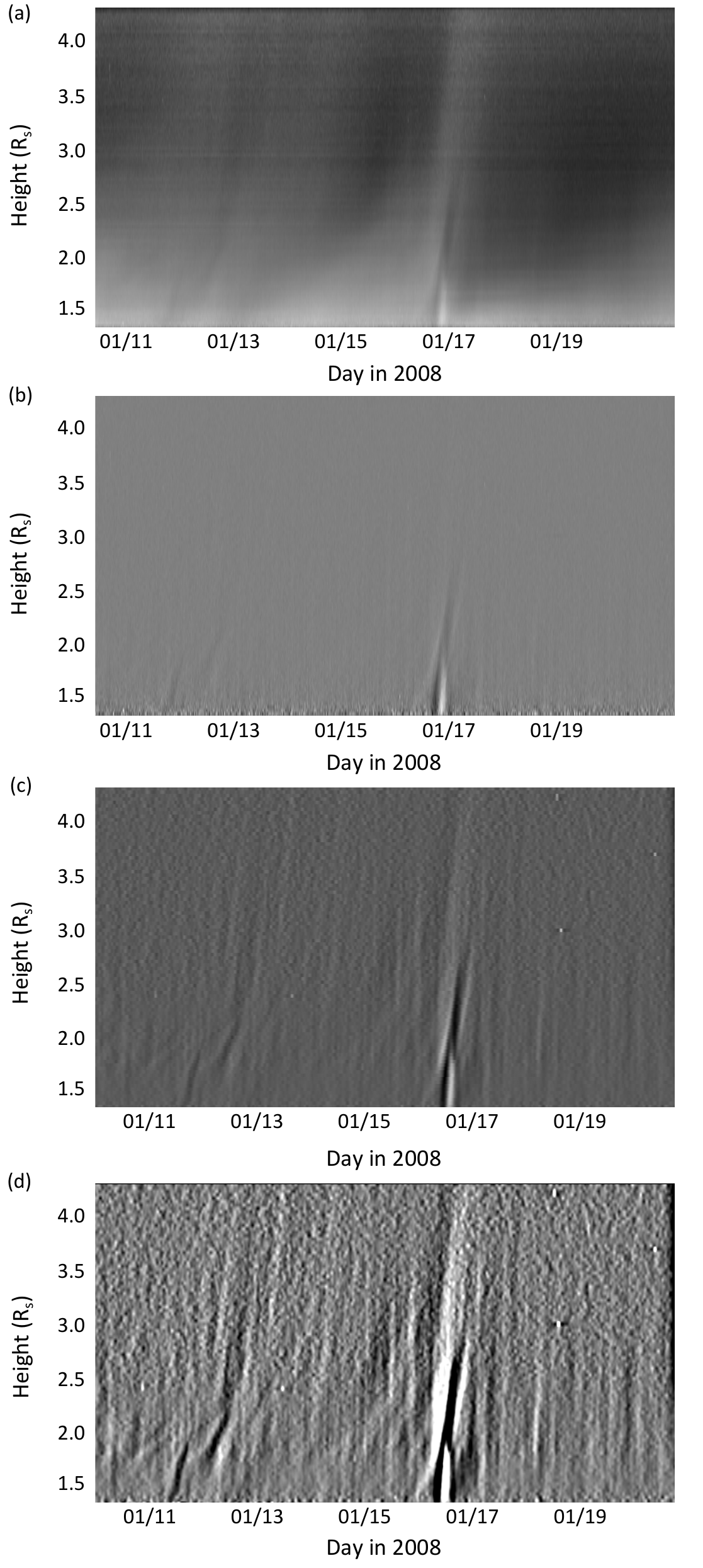}
\caption{STEREO A/COR1 Ht-T profiles for 10--21 January 2008 at various steps during the filtering process.  Each plot is for one slice of rebinned point-filtered pB data, as described in Section \ref{sec:meth-plot}, centered at PA=70$^{\circ}$, with $\Delta pa=5^{\circ}$.  (a)  Ht-T plot for slice of original time series prior to application of the bandpass filter.  (b)  Result following convolution with the wide Gaussian kernel subtracted from (a).  (c)  Result following convolution with the narrow Gaussian kernel.  (d)  Result following normalization for display.  \label{fig:bff-plot}}
\end{figure}
%%%%%%%%%%%%%%%%%%%%%%%

Figure \ref{fig:bff-plot} shows the results at various steps following the processing, for 11--20 January 2008.  Extracting a slice centered at PA=70$^{\circ}$ from the datacube results in the Ht-T plot shown in panel (a).  For display of this image, we multiplied the pixel values by the square of the height.  Only the brighter, larger-scale tracks are distinguishable.  The next step of applying the wide Gaussian kernel to the datacube and subtracting the results from (a) yields the plot shown in panel (b).  The tracks of features that change over large timescales are filtered out and what remain are tracks of small-scale features that would otherwise appear faint against the bright global structure.  Panel (c) shows the result following the convolution with the narrow Gaussian kernel to the wide-Gaussian filtered datacube.  Here, much of the noise is filtered out while the tracks of small-scale features are enhanced.  The plot shown in (d) is the result of applying a normalization of brightness at each given height in the datacube.  As such, the tracks of small-scale features are further enhanced, thus enabling tracking to larger heights.  As the method is applied onto each bin of the datacube independently, the tracks in the plots are of real features and not artifacts of the processing.  Further, the dark areas surrounding the brightest features (e.g. the tracks on the 12$^{th}$ and the 16$^{th}$) in (d) were present in (a), which shows they were not introduced during processing by the bandpass filter.

%%%%%%%%%%%%%%%%%%%%%%%%%%%
\subsection{Method for Calculating Velocities}
\label{sec:meth-vel}
%%%%%%%%%%%%%%%%%%%%%%%%%%%

Height-Time plots can be used to derive the projected velocity of each transient structure.  Assuming that a straight line on a plot corresponds to a structure moving radially with a constant velocity, the slope of the track therefore yields the velocity of the transient moving along that path.  If a track on a plot steepens, the corresponding plane of sky motion is non-radial (as in the case of most of the transients we identified in this study) and/or undergoes acceleration.

To calculate velocities, we manually divided each track into segments where the slope of the track changes and chose five points along each segment.  We made direct velocity estimates using consecutive points (time $t$, height $h$) to calculate four velocities for each segment, which we then averaged to obtain the average velocity of that segment.

%%%%%%%%%%%%%%%%%%%%%%%%%%%%%%%%%%%%%%%
%%%  		RESULTS AND ANALYSIS
%%%%%%%%%%%%%%%%%%%%%%%%%%%%%%%%%%%%%%%

\section{Data Analysis} 
\label{sec:anal-res}

%%%%%%%%%%%%%%%%%%%%%%%%%%%%%%%%
\subsection{Fast and Slow Events in COR1}
\label{sec:app-htt}
%%%%%%%%%%%%%%%%%%%%%%%%%%%%%%%%

\begin{figure*}
\centering
\includegraphics[width=0.95\textwidth]{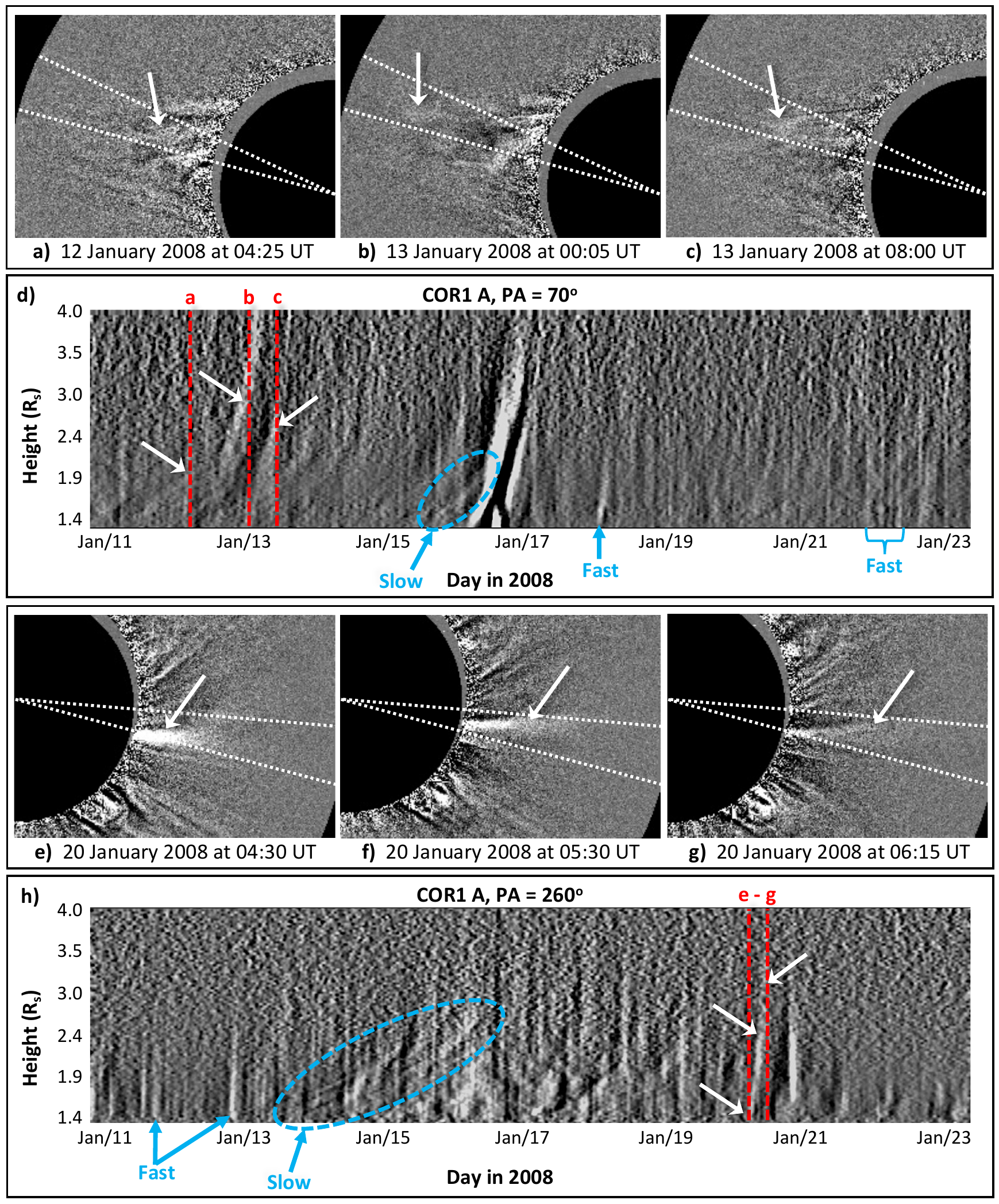}
\caption{STEREO/COR1 images and Ht-T profiles following the application of the method as described in Sections  \ref{sec:meth-img}--\ref{sec:meth-plot}.  The Ht-T profiles were generated from radial slices centered at d) PA=70$^{\circ}$ and h) PA=260$^{\circ}$, with $\Delta$pa=5$^{\circ}$.  The white dashed lines in the COR1 images trace the boundaries of each radial slice from which the plots were generated.  Red vertical dashed lines in the Ht-T plots each correspond in time to a time frame in COR1.  Examples of fast-moving structures are labeled `F' and slow-moving structures are labeled `S'.  The white arrows in the images correspond to the white arrows in the plots.  The bright feature in d) at the end of 16 January is a small `streamer-blowout' CME \citep{Sheeley_1982, Vourlidas_2018}.   \label{fig:bandpass}}
\end{figure*}

Our processed COR1 images reveal features that would otherwise remain hidden against large-scale structure and small-scale noise.  Figure \ref{fig:bandpass} shows cropped COR1 filtered images in panels a-c) and panels e-g) following the procedure as described in Section \ref{sec:meth-img} and Ht-T plots in panel d) and panel h) following the procedure as described in Section \ref{sec:meth-plot}.  The white dotted lines in the images trace the boundaries of the radial slices.

% FIGURE 5 %%%%%%%%%%%%%%%%
% Bandpass-filtered images and plots
%%%%%%%%%%%%%%%%%%%%%%%

We generated the Ht-T plots shown in panels d) and h) from a radial slice centered at PA=70$^{\circ}$ and PA=260$^{\circ}$, respectively, and spans a radial height ranging from 1.5 to 3.5 $R_{\odot}$ (from Sun center) over the 14 days of observations used for this study.  The red dashed vertical lines in panel d) (also indicated by arrows) each correspond in time to an image frame shown in panels a-c) as labeled.  The white arrows mark the point where the corresponding vertical line intersects a track on the plot, which, in turn, corresponds to the structures indicated with white arrows in panels a-c).  The red dashed vertical lines in plot h) bound the time interval during which the observations in e-g) were made.  The white arrows on the plot each denote the point along the corresponding track, which in turn corresponds, from bottom to top along the vertical, to features respectively in e), f) and g).  Note that a small streamer-blowout coronal mass ejection (CME) disrupted the east-limb streamer during January 16.  The CME has an average propagation speed of 40.8 $kms^{-1}$ and downflows associated with it can be seen in the movies provided online.

In the Ht-T plots we can distinguish a class of apparent slow flows and a class of apparent fast flows.  The tracks, which cross the red vertical lines and are marked as a, b and c in panel d), are examples of slow propagating features.  They are clear to follow and are bright compared to other tracks.  Each trajectory indicates a slow rise over heights of $\sim$1.4 to $\sim$2.2 $R_{\odot}$ during the course of a few tens of hours (up to 48 hours), giving an outward propagation speed of a few $kms^{-1}$.  The corresponding features are apparent over periods of several days and seem to trace a steeper track beyond $\sim$2.2 $R_{\odot}$, an indication that they undergo acceleration.  A subset of these slowly-propagating features, labeled `Slow' in panels d) and h), are fainter and difficult to trace beyond $\sim$2.2 $R_{\odot}$, but are also apparent over periods of several days and show coherent outward propagations.

The red lines marked `e-g' in panel h) define the short time interval over which a trajectory slopes upward, which serves as an example of a fast-moving feature.  Similar features often span the entire COR1 FOV.  They are ubiquitous throughout the 14 days of observations in this study, but vary in brightness and maximum observable height.  Assuming that they correspond to the outward motion of a single structure, they would have speeds on the order of 100 $kms^{-1}$ in the COR1 FOV.  A subset of these fast-moving features, labeled `Fast' in panel d) and h), is just above the noise level, making it difficult to distinguish the faintest ones from brighter features and noise towards the outer FOV of COR1.  However, just like their slow-moving counterparts, the fast features exhibit coherent tracks in the plots.  Figure \ref{fig:bandpass} underscores that height-time profiles are a necessary tool for the tracking of faint features identified in the images.

%%%%%%%%%%%%%%%%%%%%%%%%%%%%%%%%
\subsection{Tracking Features in EUVI, COR1 and COR2}       %
\label{sec:ht-t}    						                        %
%%%%%%%%%%%%%%%%%%%%%%%%%%%%%%%%

%%%%%%  FIGURE 6 %%%%%%%%%%%%%%
%  Tracking 
\begin{figure*}
\centering
%\begin{interactive}{animation}{cor2_filter_movie.mp4}
\centering
\includegraphics[width=\textwidth]{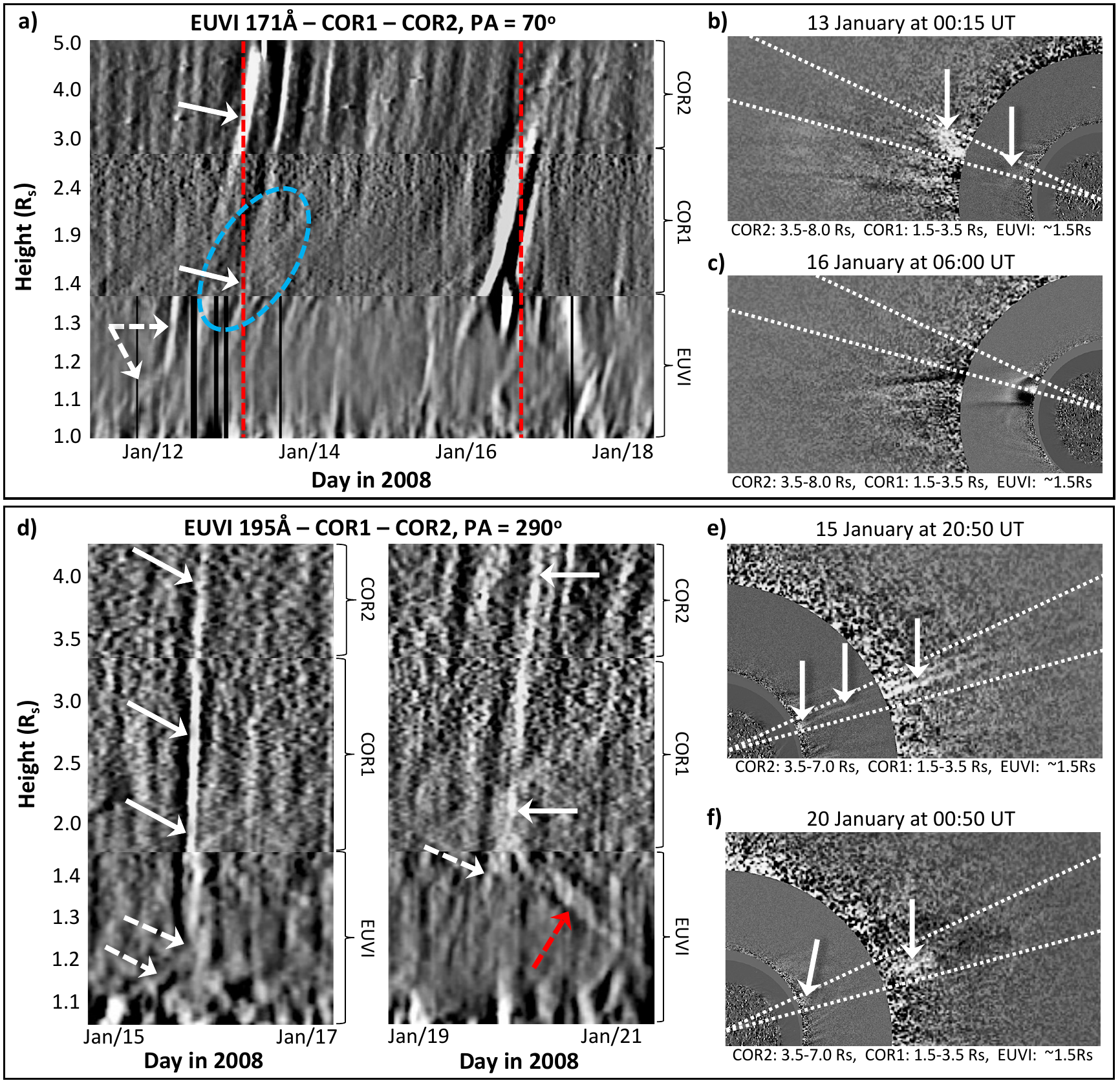}
%\end{interactive}
\caption{EUVI, COR1 and COR2 composite Ht-T profiles and EUVI, COR1 and COR2 composite images following the methods described in Sections \ref{sec:meth-img} and \ref{sec:meth-plot}.  Approximately six of the 14 days are displayed for radial slices centered at a) PA=70$^{\circ}$ and 2 days for d) PA=290$^{\circ}$, with $\Delta$pa=5$^{\circ}$.  The red vertical dashed lines in a) correspond in time to the images in b) and c), respectively.  The slant blue dotted oval in a) indicates an `X-shape' track.  The white solid arrows in a) and d) mark the points along the track(s) of features indicated in the corresponding composite images in b-c) and e-f).  The white and red dashed arrows point to tracks in EUVI.  The black vertical sections in EUVI are data gaps.  The white dotted lines in the composite images trace the boundaries of each radial slice from which the plots were generated.  A movie of processed (filtered) COR2 images spanning $\sim$12 of the 14 days (shown 11-22 January) used in this study is available online.   \label{fig:track} }
\end{figure*} 
%%%%%%%%%%%%%%%%%%%%%%%

To establish a link among structures observed by EUVI, COR1 and COR2, we generated Ht-T profile composites using all three datasets.  Figure \ref{fig:track} shows three examples where our processing facilitated the tracking of features across the three FOVs.  The height ranges displayed for the plots are shorter than the actual instrument FOV for each dataset to avoid the high level of noise near the edge of each corresponding FOV.  Additionally, since the purpose here is to track features that cross into a subsequent FOV, the height range shown for COR2 is truncated at $\sim$4.0-5.0 $R_{\odot}$.  The times shown for the corresponding EUVI, COR1 and COR2 composite images are approximate since the observation cadences are different between instruments.

Panels a) and d) show Ht-T composites within 11-21 January, along the position angle PA=70$^{\circ}$ and PA=290$^{\circ}$, respectively.  Panels b) and c) show composite images corresponding to the red vertical dashed lines in panel a).  The vertical line on the 16$^{th}$ crosses the track of the CME from Figure \ref{fig:bandpass}, which is shown here in panel c).  Panels e) and f) are the composite images at approximately the times marked by the white arrows respectively in the left and right composite plots in panel d).  The white dotted lines in the composite images trace the boundary of each radial slice from which the plot is generated.

In panel a) and b), we show examples of slow-moving features encompassing EUVI, COR1 and COR2.  Around the 12$^{th}$, a bright track in EUVI (white dashed arrows) originates in the EUV low corona and can be tracked through COR1 and COR2.  The average slope of its trajectory in EUVI corresponds to an average propagation speed of $\sim$22.5 $kms^{-1}$ between 1.2 and 1.6 $R_{\odot}$.  Between the 12$^{th}$ and the 13$^{th}$, slower tracks appear to originate near 1.25 $R_{\odot}$ in EUVI and continue into COR1 to form the `X-shape' track on the 13$^{th}$ (indicated with a slant blue dotted oval).  The average slope of their trajectories in EUVI corresponds to an average propagation speed of $\sim$12.5 $kms^{-1}$ between 1.2 and 1.6 $R_{\odot}$.  The `X-shape' track spans a radial distance from $\sim$1.4 to $\sim$2.20 $R_{\odot}$ over the course of roughly 18 hours.  Assuming the track corresponds to a single outward propagating structure, the propagation speed is $\sim$5.20 $kms^{-1}$.  However, we note that the `X-shape' may be two separate structures that appear to follow such a trajectory due to projection effects.  Regardless, the `X-shape' track exhibits a shallow slope for several hours before moving along a steeper trajectory, which is characteristic of the slow-moving structures revealed by our filtering method.

In panels d-f), we show examples of fast-moving features.  The bright track during the 15$^{th}$ has an average slope corresponding to an average propagation speed of 109 $kms^{-1}$.  This track starts from a height of $\sim$1.25 $R_{\odot}$ in the EUVI FOV (white dashed arrows mark possible starting heights).  We identified the corresponding feature in the EUVI, COR1, COR2 composite image shown in e) and indicated with white arrows.  Similarly, the steep track marked with the white arrows in COR1 and COR2 at the end of the 19$^{th}$ corresponds to the feature indicated in f) and has an average slope corresponding to an average propagation speed of 85.9 $kms^{-1}$.  Interestingly, the red dashed arrow points to a track in EUVI, which has a negative average slope suggesting an inflow with average speed of 2.0 $kms^{-1}$ between 1.45 and 1.2 $R_{\odot}$.  This inflow seems to originate from the base of the fast feature formed between 1.45 and 1.50 $R_{\odot}$.

These examples show that our methodology is successful in connecting features from the COR1 through COR2 FOV and some from the EUVI through COR2 FOV.  A movie of full disk COR2 filtered images is provided online where the evolution and propagation of additional features can be seen.

%%%%%%%%%%%%%%%%%%%%%%%%%%%%%%%%%%%
\subsection{Correlation Between Features in COR1 and COR2}       %
\label{sec:correl}    						           			%
%%%%%%%%%%%%%%%%%%%%%%%%%%%%%%%%%%%

\begin{figure}
\centering
\includegraphics[width=1.0\textwidth]{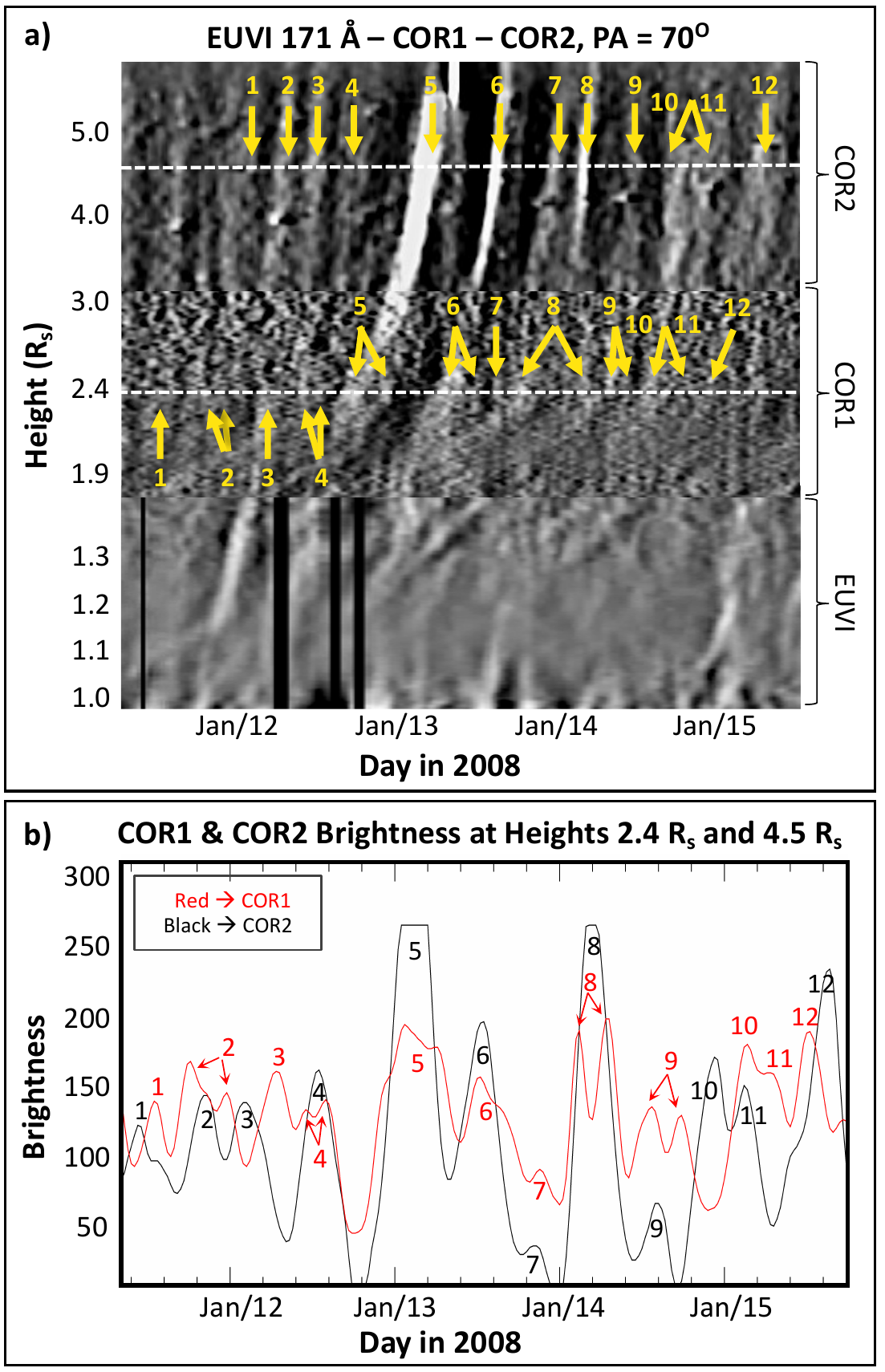}
\caption{a)  EUVI, COR1 and COR2 composite Ht-T profiles generated from a radial slice centered at PA=70$^{\circ}$, with $\Delta$pa=5$^{\circ}$, shown for 11-16 January 2008.  The white dotted horizontal lines at $Ht=2.4$ and $Ht=4.5$ $R_{\odot}$ indicate the heights at which horizontal slices were extracted to generate the plot shown in b).  b)  Brightness signal over time along the slices at heights 2.4 $R_{\odot}$ (red curve) and 4.5 $R_{\odot}$ (black curve).  Corresponding peaks are labeled.  The red curve is shown after a shift forward in time by 10 hours.   \label{fig:cross}}
\end{figure}

To provide an objective connection between features tracked in COR1 and COR2, we investigate the correlation between features in COR1 and those in COR2.  Figure \ref{fig:cross} shows the brightness signal correlation between COR1 and COR2.  Panel a) is the same Ht-T plot as shown in Figure \ref{fig:track}a, but only three days are shown here.  The white dotted horizontal lines at $Ht=2.4$ and $Ht=4.5$ $R_{\odot}$ indicate the heights at which horizontal slices were extracted to generate the plot shown in b).  The arrows in a) indicate the points at which the slices intersect the tracks on the plot (arrows are placed just above the horizontal slice to avoid covering the point of intersection).  These points are labeled with numbers, which correspond to the peaks labeled in b).  In b), the time series of brightness along the horizontal slice in COR1 is represented by the red curve and the time series of brightness along the horizontal slice in COR2 is represented by the black curve.  Red numbers mark the peaks in COR1 signal that correspond to peaks marked in COR2 signal with black numbers.

%%%%%%%%  FIGURE 7 %%%%%%%%%%%
%  Cross-correlation
%%%%%%%%%%%%%%%%%%%%%%%%%

The cross-correlation between the time series of COR1 and COR2 brightness provides the lag time of $\approx$10 hours, corresponding to the average transit time between the structures at the two different heights and obtain a delay of $\approx$10 hours.  The correlation coefficient between the two time series is $\rho=0.5$.  An estimate of the velocity at which the structures propagate between the two horizontal slices can be done using the lag time.  This gives a velocity of $\sim$41 $kms^{-1}$ for the time period shown in b).  The red curve shown is the brightness signal shifted forward by the 10 hour lag time and overlaid with the black curve.  The highest peaks clearly correspond to the brightest tracks in the COR1 and COR2 region of the plot.  On the Ht-T plot, the tracks corresponding to distinct double peaks are shown with double arrows.  The corresponding tracks for peaks 10 and 11 are also shown with double arrows because it is difficult to distinguish them as separate tracks or parts of the same track on the plot.  For peaks 4, 6, 7 and 9 we find a clear correspondence between the two instruments.  Peaks 1-3 and 10-12, however, are offset between the two instruments.  One explanation for this might be the difference in the velocities at which each corresponding feature propagates in each FOV with respect to 5 and 8, for example.

%%%%%%%%%%%%%%%%%%%%%%%%%%%%%%%%%%%%%%%
%%%  		DISCUSSION
%%%%%%%%%%%%%%%%%%%%%%%%%%%%%%%%%%%%%%%

\section{Discussion}
 \label{sec:discuss}

Our results suggest the co-existence of two types of upward propagating slow- and fast-moving structures.  Table \ref{table:transients} shows the plane-of-sky velocities estimated from our processed data in EUVI, COR1 and COR2.  In the table, we report the minimum and maximum velocity reached by a single feature as well as the average of all the observed velocities.  Based on our results, both types of features often originate low in the corona, to within tenths of a solar radius from the Sun surface.  However, their distinct properties, especially in the EUVI and COR1 FOV, suggest two separate physical mechanisms underlying them.

The slow-moving structures appear in the Ht-T plots as a series of faint recurring traces with shallow initial slopes and a gradual and continuous expansion.  They have very slow propagation at low heights (a few, or a few tens, of $kms^{-1}$), but fast in the COR2 FOV.  Looking at larger heights only, these features would be difficult to distinguish from the fast-moving structures, with an average speed at the lower range of the fast events.  It is only by combining data from COR1 and EUVI that they can be classified as a distinctly different phenomenon.

The very low speeds of the slow-moving features at low heights suggest that they are not plasma parcels carried within the solar wind outflow, but instead may be signatures of the slow evolution of the magnetic field.  The Ht-T profiles are akin to the profiles of small expanding active region loops reported by \citet{Morgan_2013}.  They report expansion speeds of $\sim$10 $kms^{-1}$ at heights up to $\sim$1.3 $R_{\odot}$, $\sim$20 $kms^{-1}$ at heights up to $\sim$2.3 $R_{\odot}$ and $\sim$60 $kms^{-1}$ at heights of $\sim$5.0 $R_{\odot}$.  The Ht-T plots in \citet{Morgan_2013} show traces as curved enhancements occurring every 3-4 hours over a period of 1-3 days, much like the enhancements in COR1 we see here in Figures \ref{fig:bandpass}d and \ref{fig:track}a.  In our profiles, the traces of slow-moving structures appear as curved enhancements every 2-4 hours and appear to last up to $\sim$2 days while undergoing acceleration at similar heights.  However, unlike the analysis of \citet{Morgan_2013}, we did not identify any active regions during the 14 days of observations used in our study.  In the absence of active regions, we may be seeing another type of small-scale dynamic activity that requires further analysis to understand.  One important note is that the bright streamer tip rotating in the in the plane of sky could exhibit apparent motions of a few $kms^{-1}$.  However, the observed acceleration by the slow features and their continuous track, observed above $\sim$ 2.5 $R_{\odot}$ and into the FOV of COR2, suggests helmet streamer tip reconnection.

The fast features at larger heights ($\geq$3.0 $R_{\odot}$) exhibit plane-of-sky velocity ranges similar to those reported for streamer blobs, or `Sheeley blobs' \citep{Sheeley_1997}.  In their analysis, \citet{Sheeley_1997} identified many events propagating at speeds that gradually accelerate from 0--200 $kms^{-1}$ near $\sim$3.0 $R_{\odot}$ to 250--300 $kms^{-1}$ near $\sim$25 $R_{\odot}$.  \citet{DeForest_2018}, as part of their work on tracking fine-scale structures in the outer corona ranging in size over multiple scales, included `Sheeley blobs' at the large-scale end of the spectrum.  They reported plane-of-sky speeds ranging from 199 to 299 $kms^{-1}$ between 4.9 and 13 $R_{\odot}$ and from 200 to 390 $kms^{-1}$ between 4.8 and 13 $R_{\odot}$.  However, looking at lower heights we can clearly notice the difference between the two types of flows.  The fast-moving structures appear to originate from a source height as low as $\sim$1.2 $R_{\odot}$ in some cases.  Such low heights have been previously reported by \citet{Jones_2009}, who tracked plasma blobs in the COR1 FOV between $\sim$1.5 and 3.5 $R_{\odot}$.  In addition, the existence of inflows at low heights suggests that at least some of the transients are formed by reconnection, possibly at the closed-field cusps of streamers (pinch-off reconnection), which is in general agreement with previous studies \citep[see][]{SheeleyWang_2007}.  \citet{Diaz_2017} determined that `Sheeley blobs' are formed via reconnection and reported blobs and inflows separating and propagating away from one another in COR2 between 5.0 and 6.0 $R_{\odot}$ during an active Sun period.  However, some of the fainter fast-moving structures very low in the corona may be due to reconnection events occurring at lower heights.  There may be reconnection events occurring at the cusp of smaller-scale loops underlying large helmet streamers, or between small loops and open-field boundaries \citep[e.g.,][]{Antiochos_2011, Drake_2020}. This may lead to a direct injection of a plasma blob that propagates outward through the main streamer body.

\begin{table}
%\centering
\setlength{\tabcolsep}{7pt}
\begin{tabular}{| c | c | c | c | c |}
\hline
 \it{STEREO}   &            Fast              &                   Fast                  &             Slow                &                  Slow                  \\ 
        Data        &         min - max         &     $\langle$ v $\rangle$     &          min - max           &     $\langle$ v $\rangle$      \\
 \hline\hline
        EUVI       &      $\sim$19 - 13      &                $\sim$29             &       $\sim$1.6 - 4.0      &              $\sim$2.5               \\
        COR1     &     $\sim$95 - 135     &               $\sim$131            &       $\sim$3.2 - 6.4      &              $\sim$4.4                \\
        COR2     &    $\sim$160 - 266    &               $\sim$208            &     $\sim$152 - 177      &             $\sim$162                \\
\hline
\end{tabular}
\caption{Plane-of-sky velocities, in $kms^{-1}$, estimated from EUVI, COR1 and COR2.  Listed are the minimum and maximum velocities and the average of all observed velocities.}
\label{table:transients}
\end{table}

We remark that the Ht-T plots were generated using radial slices, which do not capture the non-radial propagation between EUVI and COR1.  In fact, a curved track would indicate either an accelerating structure or a structure moving non-radially.  If the non-radial motion occurs in the plane of the sky, the feature will cross many radial slices and thus yield Ht-T tracks that appear and disappear transiently in neighboring slices.  This would result in a brightening and fading of the tracks that do not correspond to the true brightening and fading in the image time series.  If the non-radial motion occurs along the line of sight, the projected velocity will change, and thus the feature will (falsely) appear to speed up or slow down.  While the former can be checked by looking for the appearance of brief bright traces in Ht-T plots corresponding to neighboring position angles, the latter needs additional constraints possibly given by other observations and/or modeling of the low-middle corona.

Both the fast- and slow-moving structures are generally faint, barely above the high noise level of COR1.  They are identifiable in the Ht-T plots due to their coherent tracks.  While bandpass filtering can lead to oscillatory features, particularly at the boundaries of bright discontinuities, there are two compelling arguments to these features not being an artifact of the processing.  First, some of these features can be traced from the FOV of EUVI, through COR1, and COR2.  Given the different cadence and independent processing of the three instruments, this continuation would be highly unlikely for a method artifact.  Second, the features are not vertical stripes, they have a gradient indicative of outwards propagation.  This is an argument against the features arising from general brightness variations in the images, which would lead to vertical features.

%%%%%%%%%%%%%%%%%%%%%%%%%%%%%%%%%%%%%%%
%%%  		SUMMARY & CONCLUSIONS
%%%%%%%%%%%%%%%%%%%%%%%%%%%%%%%%%%%%%%%

\section{Conclusions}
 \label{sec:conclusions}

We present a method of using temporal bandpass filtering to process \textit{STEREO}/COR1.  This method reveals a dynamic connection between the Sun and the Heliosphere on small, sub-streamer scales.  By extending its application to COR2 and EUVI, we were able to track fast- and slow-moving propagations across the three FOVs and, in doing so, we established a continuous plane of sky view of the extended corona, thus linking the low to the high corona.  As a result, our method yields novel insight into the relationship between low coronal structures and density structures above 2.5 $R_{\odot}$.  The small-scale outward propagating structures reach a height where they must, at least partly, contribute to slow solar wind variability on timescales of hours.  Our analysis of 14 days of COR1 observations for 10-23 January 2008 led to the following results:

\begin{enumerate}

\item Our method isolated two time-varying components in COR1 data:  fast and slow outward propagating structures.

\item By tracking these structures across the different fields of view, we established a link between EUV observations of small-scale transients in the low corona and white-light observations of small-scale transients in the high corona.

\item Our method also revealed a small streamer blowout CME on 16 January 2008.  We identified associated downflows, which can be seen in the movies provided online.  This small CME has an average propagation speed of 40.8 $kms^{-1}$.  

\item Both the fast and slow transients may originate from heliocentric distances as low as $\sim$1.2 $R_{\odot}$, very low in the corona.  We suggest reconnection events within small loops contained within helmet streamers as one possible mechanism for their formation

\item The slow-moving features propagate at an average speed of 2.5 $kms^{-1}$ in EUVI, 4.4 $kms^{-1}$ in COR1 and 162 $kms^{-1}$ in COR2.  Their trajectories undergo a sharp change in slope in the Ht-T maps at $\sim$2.2 $R_{\odot}$, suggesting that acceleration occurs just below the inner edge of the COR2 FOV where converging flow (in the plane of sky) becomes radial.  This underscores the importance of COR1 observations for probing the sources of the solar wind.  

\item The fast-moving features propagate at an average speed of 29 $kms^{-1}$ in EUVI, 131 $kms^{-1}$ in COR1, and 208 $kms^{-1}$ in COR2.  These speeds fall within the range of speeds reported for the solar wind around streamers.  Similar to \citet{Jones_2009}, we report lower starting heights than previously reported---as low as $\sim$1.2 $R_{\odot}$ in some cases.  

\end{enumerate}

Our method for processing COR1 images enables a more in depth study of small-scale outward propagating plasma in the low corona by revealing faint small-scale transient structures that would otherwise be hidden.   As a result, we can follow these structures in time and identify likely solar wind source locations as they continue to release plasma structures during similar time periods.  We remark that the choice of the method parameters intrinsically determine size scales ($\Delta$ht, $\Delta$pa) and timescales of the features that can be revealed, but it does not affect the overall properties of the observed features.  The method can be modified to account for structures on other timescales.  The 14 days of observations we analyzed as part of this work was a period of quiet Sun activity in which small-scale periodic density structures (PDSs) were previously identified in \textit{STEREO A}/HI \citep{Viall_2010} and \textit{STEREO A}/COR2 \citep{Viall_2015} data in the slow solar wind.  A subsequent study will focus on re-applying our method with the filters tuned to shorter time scales to test whether the fast and slow features identified here are related to any of the shorter structures ($\sim$90 minutes) reported by \citet{Viall_2015}.

Another future effort will be to trace non-radial paths that more closely follow the underlying shapes of the helmet streamers. This is of course complicated by the fact that the streamers move and change shape with solar rotation over time periods of days or longer. Success in this will remove one source of uncertainty arising from this paper's radial-only method.

Our method  has facilitated the interpretation of Parker Solar Probe (PSP) data (PSP flies through the FOV of COR2).  It was applied to ancillary COR1 data for the study of small streamer blowout CME events observed by PSP, one in November 2018 and one in March 2019 \citep[see ][]{Nieves-Chinchilla_2020, Korreck_2020, Lario_2020}.  Given our method's flexibility and extensibility, we anticipate that this technique could be similarly applied to conjunctions with Solar Orbiter.  Our work provides a new avenue for studies of the low corona and the sources of the slow solar wind.  By combining images from remote sensing observations and \textit{in situ} measurements, the variability of the slow solar wind can be characterized, and thus expand our current knowledge of the inner heliosphere and the sources that lead to variability in the slow solar wind.

%%%%%%%%%%%%%%%%%%%%%%%%%%
%%%%%  ACKNOWLEDGMENTS  %%%%%%%%
%%%%%%%%%%%%%%%%%%%%%%%%%%

%\acknowledgments

\begin{acknowledgments}
N. Alzate acknowledges support from NASA through an appointment to the NASA Postdoctoral Program at the Goddard Space Flight Center, administered by Universities Space Research Association under contract with NASA, and support under HGI Grant No. 80NSSC20K1070.  H. Morgan acknowledges STFC grant ST/S000518/1 and Leverhulme grant RPG-2019-361 to Aberystwyth University.  N. M. Viall is supported by the NASA Heliophysics Internal Scientist Funding Model.  A. Vourlidas is supported by NASA grants 80NSSC19K1261 and NNX17AC47G.  The STEREO/SECCHI/COR2 CME catalog is generated and maintained at JHUAPL, in collaboration with the NRL and GSFC, and is supported by NASA. 
\end{acknowledgments}

%%%%%%%%%%%%%%%%
%%%%  FACILITIES  %%%%%
%%%%%%%%%%%%%%%%

%% To help institutions obtain information on the effectiveness of their 
%% telescopes the AAS Journals has created a group of keywords for telescope 
%% facilities.
%
%% Following the acknowledgments section, use the following syntax and the
%% \facility{} or \facilities{} macros to list the keywords of facilities used 
%% in the research for the paper.  Each keyword is check against the master 
%% list during copy editing.  Individual instruments can be provided in 
%% parentheses, after the keyword, but they are not verified.

\vspace{5mm}
\facilities{STEREO (EUVI, COR1, COR2)}

%%%%%%%%%%%%%%%%%%%%%%%%%%%%%%%%%%%%%%%
%%%  		REFERENCES
%%%%%%%%%%%%%%%%%%%%%%%%%%%%%%%%%%%%%%%

%\bibliography{sample63}{}

\begin{thebibliography}{}

\bibitem[Alzate \& Morgan (2016)]{Alzate_2016} Alzate, N. \& Morgan, H., 2016, \apj, \textbf{823}, 129

\bibitem[Alzate \& Morgan (2017)]{Alzate_2017} Alzate, N. \& Morgan, H., 2017, \apj, \textbf{840}, 103 

\bibitem[Antiochos et al.~(2011)]{Antiochos_2011} Antiochos, S. K., Miki\'{c}, Z., Titov, V. S., et al., 2011, \apj, \textbf{731}, 112

\bibitem[Decraemer et al.~(2019)]{Decraemer_2019} Decraemer, B., Zhukov, A. N., \& Van Doorsselaere, T., 2019, \apj, \textbf{883}, 152

\bibitem[DeForest et al.~(2018)]{DeForest_2018} DeForest, C. E., Howard, R. A., Velli, M., et al., 2018, \apj, \textbf{862}, 18

\bibitem[Drake et al.~(2020)]{Drake_2020} Drake, J. F., Agapitov, O., Swisdak, M., et al., 2020, \aap,  \url{https://doi.org/10.1051/0004-6361/202039432}

\bibitem[Freeland \& Handy (1998)]{Freeland_1998} Freeland, S. L. \& Handy, B. N., 1998, \solphys, \textbf{182}, 497

\bibitem[Gosling et al.~(1981)]{Gosling_1981} Gosling, J. T., Borrini, G., Asbridge, J. R., et al., 1981, \jgr, \textbf{86}, A7

\bibitem[Howard et al.~(2008)]{Howard_2008} Howard, R. A., Moses, J. D., Vourlidas, A., et al., 2008, \ssr, \textbf{136}, 67

\bibitem[Hundhaussen (1977)]{Hundhaussen_1977} Hundhaussen, A. J., 1977, On Coronal Holes and High Speed Wind Streams, ed. J. B. Zirker (Boulder, CO: Colorado Associated Univ. Press)

\bibitem[Jones \& Davila (2009)]{Jones_2009} Jones, S. I. \& Davila, J. M., 2009, \apj, \textbf{701}, 1906

\bibitem[Kaiser et al.~(2008)]{Kaiser_2008} Kaiser, M. L., Kucera, T. A., Davila, J. M., et al., 2008, \ssr, \textbf{136}, 5

\bibitem[Korreck et al.~(2020)]{Korreck_2020} Korreck, K. E., Szabo, A., Nieves-Chinchilla, T., et al., 2020, \apjs, \textbf{246}, 69

\bibitem[Lario et al.~(2020)]{Lario_2020} Lario, D., Balmaceda, L., Alzate, N., et al., 2020, \apj, \textbf{897}, 134

\bibitem[McComas et al.~(1998)]{McComas_1998} McComas, D. J., Barne, S. J., Barraclough, B. L., et al., 1998, \grl \textbf{25}, 1
   
\bibitem[Morgan et al.~(2006)]{Morgan_2006} Morgan, H., Habbal, S. R., \& Woo, R., 2006, \solphys, \textbf{236}, 263
   
\bibitem[Morgan et al.~(2012)]{Morgan_2012} Morgan, H., Byrne, J. P., \& Habbal, S. R., 2012, \apj, \textbf{752}, 144

\bibitem[Morgan et al.~(2013)]{Morgan_2013} Morgan, H., Jeska, L., \& Leonard, D., 2013, \apjs, \textbf{206}, 19
   
\bibitem[Morgan \& Druckm\"{u}ller (2014)]{Morgan_2014} Morgan, H., \& Druckm\"{u}ller, M., 2014, \solphys, \textbf{289}, 2945

\bibitem[Morgan \& Cook (2020)]{Morgan_2020} Morgan, H., \& Cook, A. C., 2020, \apj, \textbf{893}, 57

\bibitem[Nieves-Chinchilla et al.~(2020)]{Nieves-Chinchilla_2020} Nieves-Chinchilla, T., Szabo, A., Korreck, K. E., et al., 2020, \apjs, \textbf{246}, 63

\bibitem[S\'{a}nchez-D\'{i}az et al.~(2017)]{Diaz_2017} S\'{a}nchez-D\'{i}az, E., Rouillard, A. P., Davies, J. A., et al., 2017, \apj, \textbf{835}, L7

\bibitem[Sheeley et al.~(1982)]{Sheeley_1982} Sheeley, N. R., Howard, R. A., Koomen, M. J., et al., 1982, \ssr, \textbf{33}, 219

\bibitem[Sheeley et al.~(1997)]{Sheeley_1997} Sheeley, N. R., Wang, Y.-M., Hawley, S. H., et al., 1997, \apj, \textbf{484}, 472-478

\bibitem[Sheeley \& Wang (2007)]{SheeleyWang_2007} Sheeley, N. R. \& Wang, Y.-M., 2007, \apj, \textbf{655}, 1142-1156

\bibitem[Thompson et al.~(2003)]{Thompson_2003} Thompson, W. T., Davila, J. M., Fisher, R. R., et al., 2003, \procspie, \textbf{4853}, 1-11

\bibitem[Thompson et al.~(2010)]{Thompson_2010} Thompson, W. T., Wei, K., Burkepile, J. T., et al., 2010, \solphys, \textbf{262}, 213-231

\bibitem[Viall et al.~(2010)]{Viall_2010} Viall, N. M., Spence, H. E., Vourlidas, A., \& Howard, R. A., 2010, \solphys, \textbf{261}, 175
   
\bibitem[Viall \& Vourlidas (2015)]{Viall_2015} Viall, N. M., \& Vourlidas, A., 2015, \apj, \textbf{807}, 176

\bibitem[Vourlidas \& Webb (2018)]{Vourlidas_2018} Vourlidas, A., \& Webb, D. F., 2018, \apj, \textbf{861}, 103

\end{thebibliography}
%\bibliographystyle{aasjournal}

\end{document}